\documentclass[a4paper,11pt]{article}
\pdfoutput=1 

\usepackage{jcappub} 
\graphicspath{{plots/}}

\usepackage{lineno,hyperref}
\modulolinenumbers[5]

\usepackage{soul}
\usepackage{amsmath}
\usepackage{amssymb}
\usepackage{color}
\usepackage{xcolor}
\usepackage{todonotes}
\usepackage{upgreek}
\usepackage[normalem]{ulem}

\newcommand{\Neff}{\ensuremath{N_{\rm eff}}}
\newcommand{\K}{{\rm K}}

\title{Non-unitary three-neutrino mixing in the early Universe}

\author[a]{Stefano Gariazzo,}
\emailAdd{gariazzo@to.infn.it}
\author[b, c]{Pablo Mart\'{\i}nez-Mirav\'e,}
\emailAdd{pamarmi@ific.uv.es}
\author[b]{Olga Mena,}
\emailAdd{omena@ific.uv.es}
\author[b]{Sergio Pastor}
\emailAdd{pastor@ific.uv.es}
\author[b, c]{and Mariam T\'ortola}
\emailAdd{mariam@ific.uv.es}

\affiliation[a]{INFN, Sezione di Torino, Via P. Giuria 1, I--10125 Torino, Italy}
\affiliation[b]{Instituto de F{\'\i}sica Corpuscular  (CSIC-Universitat de Val{\`e}ncia)\\ 
	Parc Cient{\'\i}fic UV, C/ Catedr{\'a}tico Jos{\'e} Beltr{\'a}n, 2, 46980 Paterna, Spain}
\affiliation[c]{Departament de F{\'\i}sica Te\`orica, Universitat de Val{\`e}ncia, 46100 Burjassot, Spain}
\date{}

\abstract{Deviations from unitarity in the  three-neutrino mixing canonical picture are expected in many physics scenarios beyond the Standard Model. The mixing of new heavy neutral leptons with the three light neutrinos would in principle modify the strength and flavour structure of charged-current and neutral-current interactions with matter.
Non-unitarity effects would therefore have an impact on the neutrino decoupling processes in the early Universe and on the value of the effective number of neutrinos, \Neff. We calculate the cosmological energy density in the form of radiation with a non-unitary neutrino mixing matrix, addressing the possible interplay between parameters. Highly accurate measurements of \Neff\ from forthcoming cosmological observations can provide independent and complementary limits on the departures from unitarity. For completeness, we relate the scenario of small deviations from unitarity to non-standard neutrino interactions and compare the forecasted constraints to other existing limits in the literature.
}
\begin{document}
\maketitle
\flushbottom

\section{Introduction}

In recent years, an unprecedented accuracy has been reached in the extraction of the neutrino mixing parameters~\cite{deSalas:2020pgw,Esteban:2020cvm,Capozzi:2021fjo}: solar, atmospheric, reactor and accelerator neutrino oscillation experiments have  provided conclusive evidence for neutrino masses in nature. However, the underlying mechanism for generating the neutrino masses remains unknown. Although there is a plethora of possible theoretical frameworks aiming to settle this issue, generally, a common mechanism consists on the generation of the neutrino masses  through the mediation of heavy neutral leptons (HNLs). Then, the lepton mixing matrix has to account for the possible admixture between light neutrinos and the HNLs.
In seesaw models \cite{Minkowski:1977sc,Yanagida:1979as,Schechter:1980gr,Schechter:1981cv,Mohapatra:1980yp}, the smallness of neutrino masses would indicate a very high scale for the conventional realisations of this mechanism, leading to a negligible mixing between light neutrinos and HNLs \cite{Grimus:2000vj}, rendering a mixing matrix very close to unitary.
Nonetheless, in `low scale' realisations of the seesaw mechanism \cite{Mohapatra:1986bd, Akhmedov:1995ip,Akhmedov:1995vm,Malinsky:2005bi,Malinsky:2009df, Malinsky:2009gw}, much more relevant departures from unitarity can be expected.
In this context, two general scenarios can be envisioned: (i) not-so-heavy neutral leptons, with masses such that they can be produced in oscillation experiments, or (ii) canonical HNLs, which are kinematically inaccessible in oscillation experiments.
The former is being scrutinised due to its potential to explain short-baseline neutrino oscillation anomalies (see e.g.~\cite{Gariazzo:2015rra,Giunti:2019aiy,Palazzo:2020tye,Schoppmann:2021ywi,Archidiacono:2022ich}), and  its cosmological effects have been studied extensively in the past~\cite{Hannestad:2012ky,Saviano:2013ktj,Gariazzo:2019gyi,Hagstotz:2020ukm}.
However, when discussing non-unitarity (NU) of the three-neutrino mixing matrix, we shall consider exclusively the latter.
Direct searches for HNLs and precision flavour observables \cite{Antusch:2014woa} allow to constrain the light-heavy mixing for a wide range of masses. These measurements rely on the fact that the charged-current leptonic interactions are modified. For neutrino experiments operating at energies below the mass of the HNLs, neutrino neutral-current interactions are also altered.
As a consequence, complementary limits from both the non-unitarity mixing and the changes in the structure of the interactions can also be obtained within this context. 

In the early Universe, electroweak processes kept neutrinos in thermal contact with electrons, positrons and photons. Once the collisions became ineffective at temperatures around the MeV, neutrinos decoupled from the cosmic plasma. Consequently, any modification of neutrino interactions would change the picture of neutrino decoupling and alter the resulting background of relic neutrinos. 
The role of neutrinos is encoded in the effective number of relativistic species \Neff, which quantifies relativistic degrees of freedom different from photons that contribute to the cosmological radiation density. Within the canonical picture of neutrino decoupling, $\Neff = 3.0440$~\cite{Akita:2020szl,Froustey:2020mcq,Bennett:2020zkv}, in excellent agreement with the most recent determination from the combination of several cosmological probes,  $\Neff = 2.99^{+0.34}_{-0.33}$ at 95\% confidence level (CL) from Planck 2018 (TT, TE, EE + lowE + lensing + BAO)~\cite{Planck:2018vyg}.
This measurement is expected to improve significantly with forthcoming data from Cosmic Microwave Background (CMB) measurements from the Simons Observatory ($\sigma(\Neff) \simeq 0.05 - 0.07$) \cite{SimonsObservatory:2018koc} and CMB-S4 ($\sigma(\Neff ) \simeq 0.02 - 0.03$) \cite{CMB-S4:2016ple}.

In this work, we address the impact of a non-unitary three-neutrino mixing on the process of neutrino decoupling and determine the expected complementary bounds on departures from unitarity that will be obtained by near future CMB surveys via their measurement of \Neff. 
The structure of this manuscript is as follows. In Section \ref{sec:nu}, we introduce the formalism and notation adopted for the study of non-unitarity three-neutrino mixing and in Section \ref{sec:cosmo} we discuss the main implications for neutrino cosmology. The obtained results are presented in Sections \ref{sec:results} and \ref{sec:discussion}, where we also discuss a mapping to the non-standard interactions  scenario in case of small departures from unitarity, as well as the complementarity with other searches. We conclude in Section \ref{sec:conclusion}.

\section{Non-unitarity in the three-neutrino mixing scenario}
\label{sec:nu}
Let us denote the $n$-vector  representing the neutral lepton states with definite mass, both light and heavy, by $\upnu_L$. In this basis, the charged- and neutral-current (CC and NC, respectively) Lagrangians read
\begin{align}
    \mathcal{L}_ {\rm CC} = \frac{g}{\sqrt{2}} (W^-)_\mu \bar{e}_L\gamma^\mu \K \upnu_L + {\rm H.c.} \quad {\rm and} \quad \mathcal{L}_{\rm NC} = \frac{g}{2 \cos \theta_W} Z_\mu \bar{\upnu}_L (\K^\dagger \K) \upnu_L +{\rm H.c. }\, ,
\end{align}
where we have introduced the matrix \K,~which relates the three active flavour eigenstates and the $n$ mass eigenstates.\footnote{Note that the $3\times n$ matrix \K~also includes the rotation required to define the charged-leptons mass basis. } At low energies, it is possible to express the CC and NC interactions in terms of four-fermion interactions, i.e.
\begin{align}
    \mathcal{L}_{\rm CC} =-2\sqrt{2}G_{\rm F}\sum_{i,j}(\K^\dagger)_{i e}\K_{e j}\left(\bar{\nu}_i \gamma^\mu P_L \nu_j\right)\left(\bar{e}\gamma_\mu P_L e \right)~,
    \label{eqn:nucc}
\end{align}
and 
\begin{align}
    \mathcal{L}_{\rm NC} = -2\sqrt{2}G_{\rm F}\sum_{X = L, R}g_X\sum_{i,j}\left(\K^\dagger \K \right)_{ij}\left(\bar{\nu}_i \gamma^\mu P_L \nu_j\right)\left(\bar{e}\gamma_\mu P_X e \right) \, ,
    \label{eqn:nunc}
\end{align}
where in the case of Eq.~\eqref{eqn:nucc} we have performed a Fierz transformation. Here $G_{\rm F}$ is the Fermi constant and the index $X=\{L, R\}$, so that $P_X$ denotes the chiral projectors $P_{R, L}=(1\pm\gamma_5)/2$.

Note that, when computing the amplitude of the decay and scattering processes, the sum over mass eigenstates should be limited to the largest mass eigenstate kinematically accessible.
Thus, it is useful to separate the matrix \K~in two blocks,
\begin{align}
    \K = \begin{pmatrix}
        N & S 
    \end{pmatrix}\, ,
\end{align}
where $N$ describes the mixing among the three lightest states and $S$ accounts for the mixing between the three lightest states and the $n-3$ heavy ones.

In what follows, we shall assume that only three mass eigenstates lie below the energy scale of interest here. In that case, the sums in Eqs.\ (\ref{eqn:nucc})-(\ref{eqn:nunc}) only include the three lightest mass eigenstates, and only the $3\times3$ submatrix $N$ is involved. Notice that the unitarity of the full $n\times n$ lepton mixing matrix does not require the submatrix $N$ appearing in the effective CC and NC Lagrangians to be unitary. In order to account for departures of non-unitarity in $N$, we adopt the following parametrisation \cite{Escrihuela:2015wra}
\begin{equation}
    N = \begin{pmatrix}
    \alpha_{11} & 0 & 0 \\ \alpha_{21} & \alpha_{22} & 0 \\ \alpha_{31} & \alpha_{32} & \alpha_{33} 
    \end{pmatrix} U\, ,
    \label{eq:parametrisation}
\end{equation}
where $U$ is the usual unitary three-neutrino mixing matrix, $\alpha_{ii}$ are real parameters, and $\alpha_{ij} ~ (i \neq j)$ are complex and provide additional sources of CP violation~\footnote{The relation between the $\alpha_{ij}$ parameters and the parametrisation of the full $n\times n$ mixing matrix is detailed in Appendix \ref{sec:mixing}.}. The off-diagonal parameters are related to the diagonal ones through the triangular inequality~\cite{Escrihuela:2016ube},
\begin{equation}
    |\alpha_{ij}| \leq \sqrt{(1- \alpha^2_{ii})(1-\alpha^2_{jj})}\, .
\end{equation}
Additionally, from the unitarity of the full $n \times n$ mixing matrix~\cite{Forero:2021azc}, the following conditions apply
\begin{align}
    \alpha^2_{11}|\alpha_{21}|^2 &\leq (1 -\alpha^2_{11})(1-\alpha^2_{22} - |\alpha_{21}|^2) \, , \label{eq:NUcond1}\\
    \label{eq:NUcond2}
    \alpha^2_{11} |\alpha_{31}|^2 &\leq (1 -\alpha^2_{11})(1-\alpha^2_{33} - |\alpha_{31}|^2-|\alpha_{32}|^2 ) \, , \\
    |\alpha_{22}\alpha_{32} + \alpha^*_{21}\alpha_{31} |^2 &\leq (1 -\alpha^2_{22}-|\alpha_{21}|^2)(1-\alpha^2_{33} - |\alpha_{31}|^2 -|\alpha_{32}|^2)\, .
    \label{eq:NUcond3}
\end{align}

Non-unitary three-neutrino mixing would not only alter CC and NC interactions, but also the picture of neutrino oscillations, whose observations at both short- and long-baseline experiments can be used to constrain the non-unitarity parameters.
Current bounds are summarized in Table \ref{table:bounds}, from \cite{Forero:2021azc}.

\begin{table}[h!]
\centering
\begin{tabular}{|l|c|c|c|c|c|c|}
\hline
NU parameters & $|\alpha_{21}|$ & $|\alpha_{31}|$ & $|\alpha_{32}|$  & $\alpha_{11}$ & $\alpha_{22}$ & $\alpha_{33}$ \\ 
\hline
3$\sigma$ bounds   & $< 0.025$ & $< 0.075$ & $<  0.02$  & $>  0.93$ & $>  0.98$ & $>  0.72$ \\
\hline
\end{tabular}
\caption{%
Current constraints on the non-unitary parameters from Ref.~\cite{Forero:2021azc}.}
\label{table:bounds}
\end{table}

\section{The impact of non-unitary three-neutrino mixing in neutrino decoupling}
\label{sec:cosmo}

In the following, we shall concentrate in a scenario in which HNLs  can not be produced at temperatures $\sim$MeV and, therefore, their distribution is Boltzmann-suppressed. Consequently, only the evolution of the three lightest mass eigenstates needs to be studied. In previous studies of neutrino decoupling, the evolution of the neutrino distributions was mostly addressed in terms of flavour states \cite{Froustey:2020mcq,Bennett:2020zkv}, but an adequate treatment of non-unitarity requires to work in the mass basis  (see e.g.\ \cite{Akita:2020szl}). The reason is that the truncation in the sum over accessible states is only well defined for mass eigenstates.

We adopt the usual definitions for the comoving variables $x = m_e \,a$, $y = p\,a$, $z = T_\gamma\,a$, in terms of the electron mass $m_e$, the cosmological scale factor $a$, the neutrino momentum $p$, and the photon temperature $T_\gamma$
\cite{Gariazzo:2019gyi,Bennett:2020zkv}.
The evolution of the density matrix $\varrho$ for tree neutrinos in the mass basis reads \cite{Akita:2020szl}
\begin{equation}
\label{eq:drho_dx_nxn}
\frac{{\rm d}\varrho(y)}{{\rm d}x}
=
\sqrt{\frac{3 m^2_{\rm Pl}}{8\pi\rho}}
\left\{
    -i \frac{x^2}{m_e^3}
    \left[
        \frac{\mathbb{M}}{2y}
        -
        \frac{2\sqrt{2}G_{\rm F} y m_e^6}{x^6}
        \left(\frac{\mathbb{E}_\ell+\mathbb{P}_\ell}{m_W^2}\right)
            ,
    \varrho
    \right]
    +\frac{m_e^3}{x^4}\mathcal{I(\varrho)}
\right\}\,,
\end{equation}
where $m_{\rm Pl}$ denotes the Planck mass, $\rho$ is the energy density of the Universe, $m_W$ is the mass of the $W$ gauge bosons and we have defined the diagonal neutrino mass matrix $\mathbb{M} = {\rm diag}(0, \Delta m^2_{21},  \Delta m^2 _{31})$. 
We have also introduced the energy density of charged leptons, $\mathbb{E}_\ell$, and the pressure of charged leptons, $\mathbb{P}_\ell$.
In this scenario, it is useful to define the following matrices
\begin{align}
    \label{eqn:YL}
    \left(Y_L\right)_{ij} &= g_L \left(N^\dagger N\right)_{ij} + (N^\dagger)_{i e} N_{ej}\, , \\
    \label{eqn:YR}
    \left(Y_R\right)_{ij} &= g_R \left(N^\dagger N\right)_{ij}\, ,
\end{align}
with the $i$ and $j$ indices running from 1 to 3, such that the charged-lepton energy density and pressure,
\begin{align}
     \mathbb{E}_{\ell} = \rho_e \, \mathbb{E}_{\rm NU}~, & & \mathbb{P}_{\ell} = P_e\, \mathbb{E}_{\rm NU}   \simeq  \rho_e \, \mathbb{E}_{\rm NU}/3~,
\end{align}
are expressed in terms of the electron energy density and the pressure of charged leptons ($\rho_e$ and $P_e$, respectively) and also via a matrix accounting for the contribution of charged leptons to the matter potential
\begin{equation}
    \mathbb{E}_{\rm NU} = Y_L  - Y_R \, .
\end{equation}

Finally, the term $\mathcal{I(\varrho)}$ in \eqref{eq:drho_dx_nxn} encodes the collision integrals, arising from the neutrino-electron scattering and annihilation processes, as well as from the neutrino-neutrino interactions.
The terms accounting for neutrino-electron interactions are proportional to $G_{\rm F}^2$ and also depend on the matrices $Y_L$ and $Y_R$ defined in Eqs.\ (\ref{eqn:YL})-(\ref{eqn:YR}), see Appendix \ref{sec:collison} for details.
Neutrino-neutrino interactions are also modified if the three-neutrino mixing is not unitary. The low-energy Lagrangian accounting for neutrino self-interactions reads
\begin{equation}
    \mathcal{L}_{\nu SI} = -2 \sqrt{2}G_{\rm F} \sum_{i,j,k,m} \left(N^\dagger N\right)_{ij}\left(N^\dagger N\right)_{mk}(\bar{\nu}_i \gamma^\mu P_L \nu_j)(\bar{\nu}_k \gamma_\mu P_L \nu_m)\, .
\end{equation}
It has been shown that neutrino self-interactions lead to a refractive term in the Hamiltonian, which does not modify the picture qualitatively and it only gives a sub-leading contribution to \Neff~\footnote{%
The role of neutrino self-interactions in cosmology has been studied extensively and was proposed as a solution to the so-called Hubble tension, see e.g. Ref.~\cite{DiValentino:2021izs} and references therein. Note, however, that self-interactions invoked in those scenarios are orders of magnitude larger than the one expected to arise from the non-unitarity of the three-neutrino mixing.}, hence we ignore such terms.

Notice that we have treated the Fermi constant $G_{\rm F}$ as a fundamental quantity that we can precisely measure experimentally. However, in the presence of non-unitary three-neutrino mixing, what it is measured is an effective parameter depending on the mixing between light neutrinos and HNLs \cite{Escrihuela:2015wra}. The effective parameter measured in the case of $\beta$-decay is
\begin{equation}
    G^\beta_{\rm F} = G_{\rm F} \sqrt{(NN^ \dagger)_{ee}} = G_{\rm F} \alpha_{11}~,
\end{equation}
whereas for muon decay it is
\begin{equation}
    G^\mu_{\rm F}=G_{\rm F} \sqrt{(NN^\dagger)_{ee} (N N^\dagger)_{\mu\mu}} = G_{\rm{F}} \sqrt{\alpha^2_{11}(\alpha^2_{22}+ |\alpha_{21}|^2)} \, .
    \label{eq:GFmu}
\end{equation}
The measurement from muon decay has achieved a higher precision and, consequently, we take it as a reference, and consider $G^\mu_{\rm F} = 1.1663787(6)\times 10^{-5} \, {\rm GeV}^{-2}$ \cite{RevModPhys.93.025010}.

\section{Results}
\label{sec:results}

We make use in what follows of the publicly available code  \texttt{FortEPiaNO} (Fortran Evolved Primordial Neutrino Oscillations) \cite{Gariazzo:2019gyi,Bennett:2020zkv}, which has been modified in order to work in the mass basis and to include the effects from a possible non-unitary three-neutrino mixing. The numerical settings adopted here allow to compute the value of \Neff\ with a precision better than $10^{-3}$, which is safely smaller than the expected sensitivity from future surveys and, therefore, accurate enough for the study of non-unitary three-neutrino mixing.

\subsection{The role of diagonal parameters $\mathbf{\alpha_{ii}}$}

According to the parametrisation in Eq.~\eqref{eq:parametrisation}, non-unitarity could manifest as a deviation from unity in the parameters $\alpha_{ii}$ (see Appendix \ref{sec:mixing} for the relation with the parametrisation of the $n\times n$ unitary mixing matrix). To understand their impact, we address the role of each one individually, as well as the interplay between them.

\begin{figure}[t!]
    \centering
    \includegraphics[width =\textwidth]{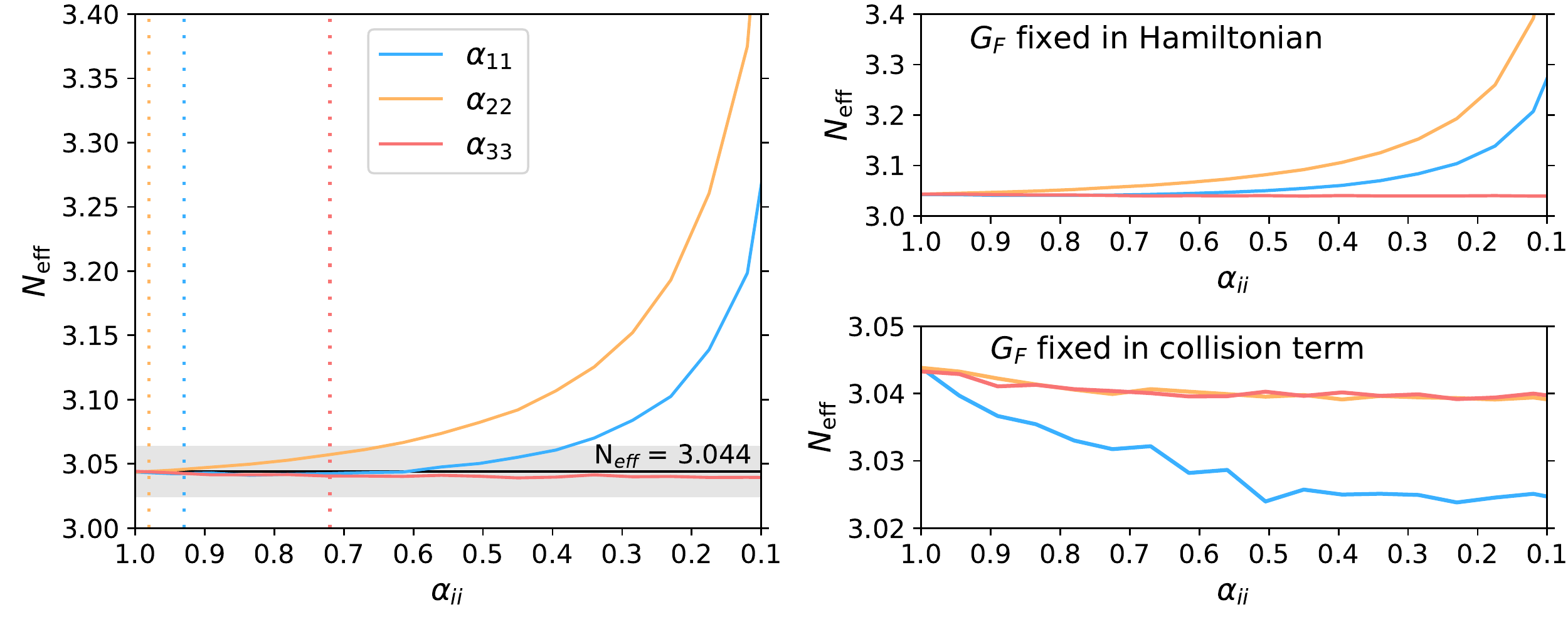}
    \caption{Values of \Neff\ as a function of  the diagonal non-unitarity parameters $\alpha_{ii}$, when only one of them is allowed to be non-zero.  For comparison, we show the exact calculation (left panel), where the vertical dotted lines indicate the existing limits (see Table \ref{table:bounds}), and the cases when the dependence of $G_{\rm F}$ on the $\alpha_{ii}$ parameters is considered only in the collision integrals (upper right panel) or in the Hamiltonian (lower right panel).}
    \label{fig:one_at_a_time}
\end{figure}

Figure \ref{fig:one_at_a_time} depicts the dependence of \Neff\ on each of the individual diagonal parameters, $\alpha_{ii}$. In the left panel, we show the exact calculation of \Neff\ as a function of $\alpha_{ii}$, allowing only one of them being different from zero simultaneously. It is known that, in other non-standard scenarios, the main contribution of new physics to \Neff\ comes from the modification of the collision integrals. For illustrative purposes, the top right panel shows the values of \Neff\ after considering the dependence of $G_{\rm F}$ on each of the $\alpha_{ii}$ parameters only in the collision terms (and not in the Hamiltonian) whereas the bottom right panel displays the opposite situation (note the different scale for \Neff). 

Notice that, since the relation between the true value of $G_{\rm F}$ and the one measured from muon decay does not depend on $\alpha_{33}$, the calculated \Neff\ is almost insensitive to this parameter.
Indeed, the main impact of non-unitarity on neutrino decoupling arises from the mismatch between the true and the observed value of $G_{\rm F}$.
The second key point is that, as it can be noticed from the right plots of Figure \ref{fig:one_at_a_time}, the collision terms dominate the dependence of \Neff\ on the non-unitarity parameters: changes on the effective $G_{\rm F}$ alter both the scattering and annihilation processes involving neutrinos, electrons and positrons, leading to a delayed decoupling and hence to a larger value of \Neff. Albeit matter effects also depend on the effective $G_{\rm F}$, this is not as relevant for neutrino decoupling as the interactions with the cosmic plasma. From the current measurement of $\Neff=2.99^{+0.34}_{-0.33}$ at 95$\%$~C.L. \cite{Planck:2018vyg}, it is possible to set weak limits on the non-unitarity of the three-neutrino mixing matrix,
\begin{align}
\centering
    & \alpha_{11} > 0.07 \, ,& & \alpha_{22}> 0.15 \, ,
\end{align}
both at $95\%$~C.L. As an exercise, we predict the improvement on these limits from future CMB observations. Let us assume a forecasted sensitivity to \Neff\ with an uncertainty $\sigma = 0.02$ \cite{CMB-S4:2016ple}, together with a Gaussian posterior distribution for this parameter. Then, one can define the $\chi^2$ test:
\begin{align}
    \chi^2 = \frac{[(N_{\rm eff})_0 - N_{\rm eff}]^2}{\sigma^2} \, ,
    \label{eq:chi2}
\end{align}
where $(\Neff)_0$ corresponds to the measured value of \Neff, assumed to match the standard theoretical prediction, i.e.\ $(\Neff)_0 = 3.044$.
Under these assumptions, we find that future CMB measurements will improve considerably the former $95\%$~C.L. upper limits:
\begin{align}
\centering
    & \alpha_{11} > 0.29 \,,& & \alpha_{22}> 0.50\, .
\end{align}

In the following, we shall present an exhaustive study of the parameter correlations. Given the $\chi^2$ test definition in Eq.~(\ref{eq:chi2}), one can extract the confidence regions expected when two parameters are allowed to differ from unity simultaneously in the analyses.
\begin{figure}
    
    \includegraphics[width = 0.48\textwidth]{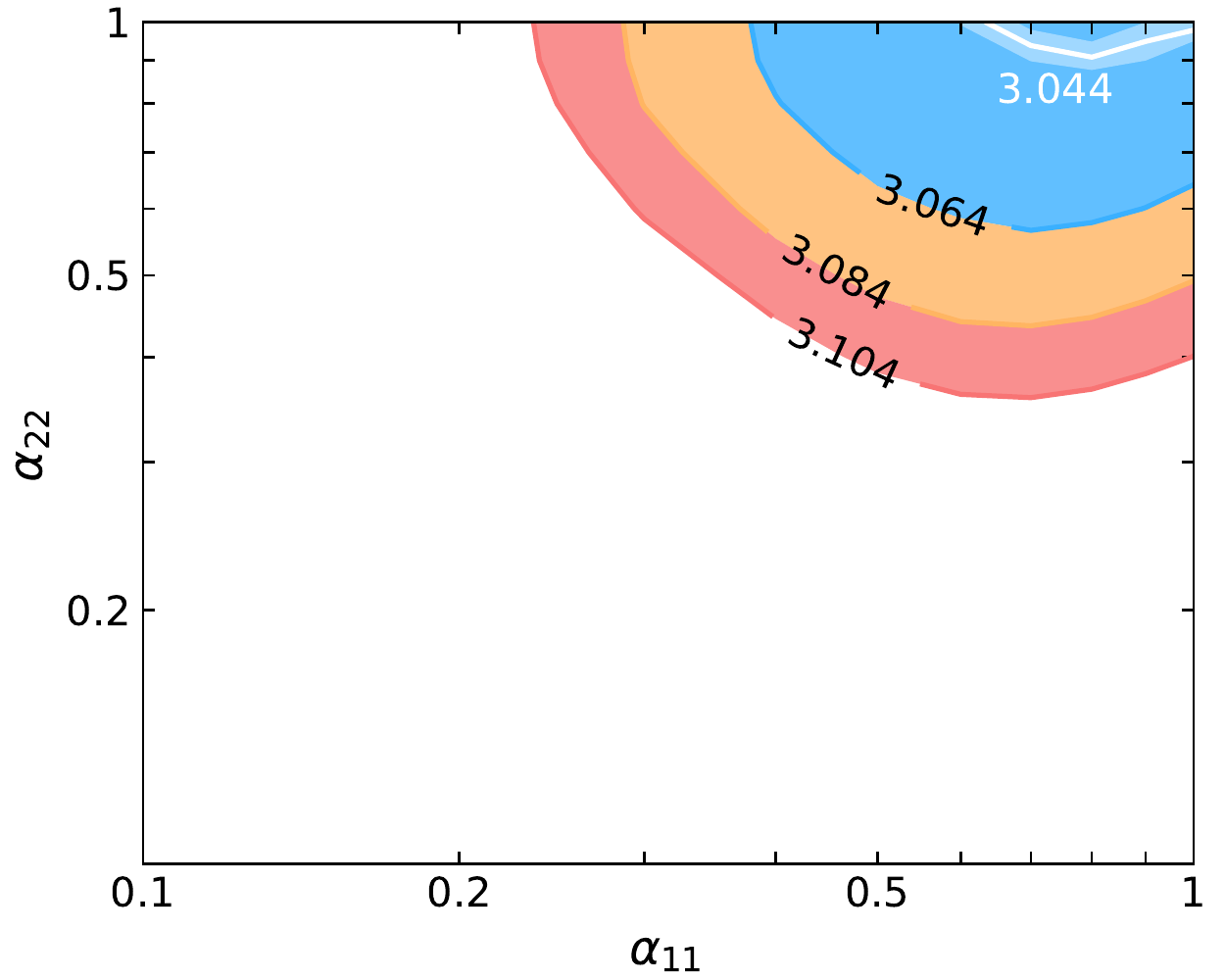}
    \\
    \includegraphics[width = 0.48\textwidth]{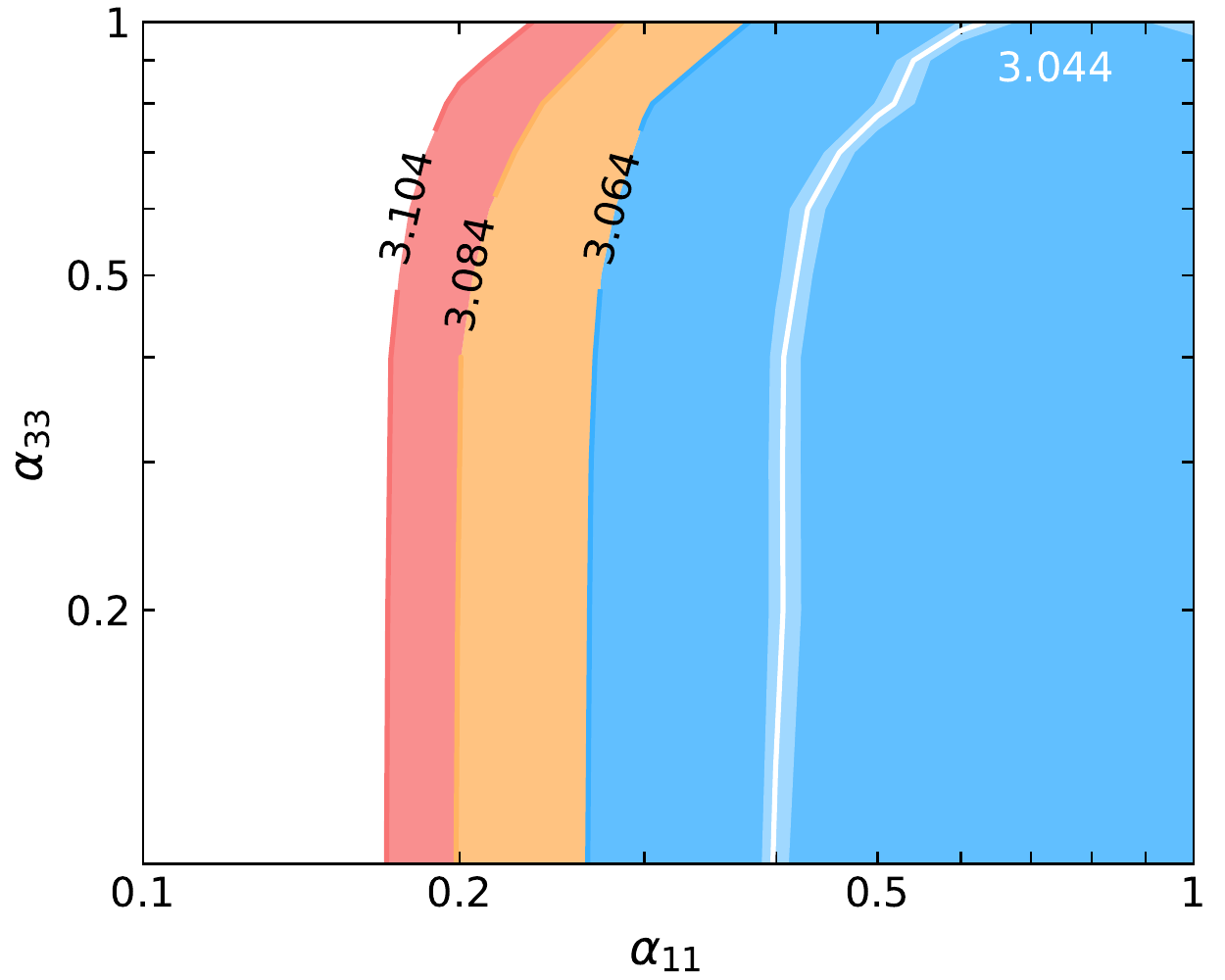}
    \includegraphics[width = 0.48\textwidth]{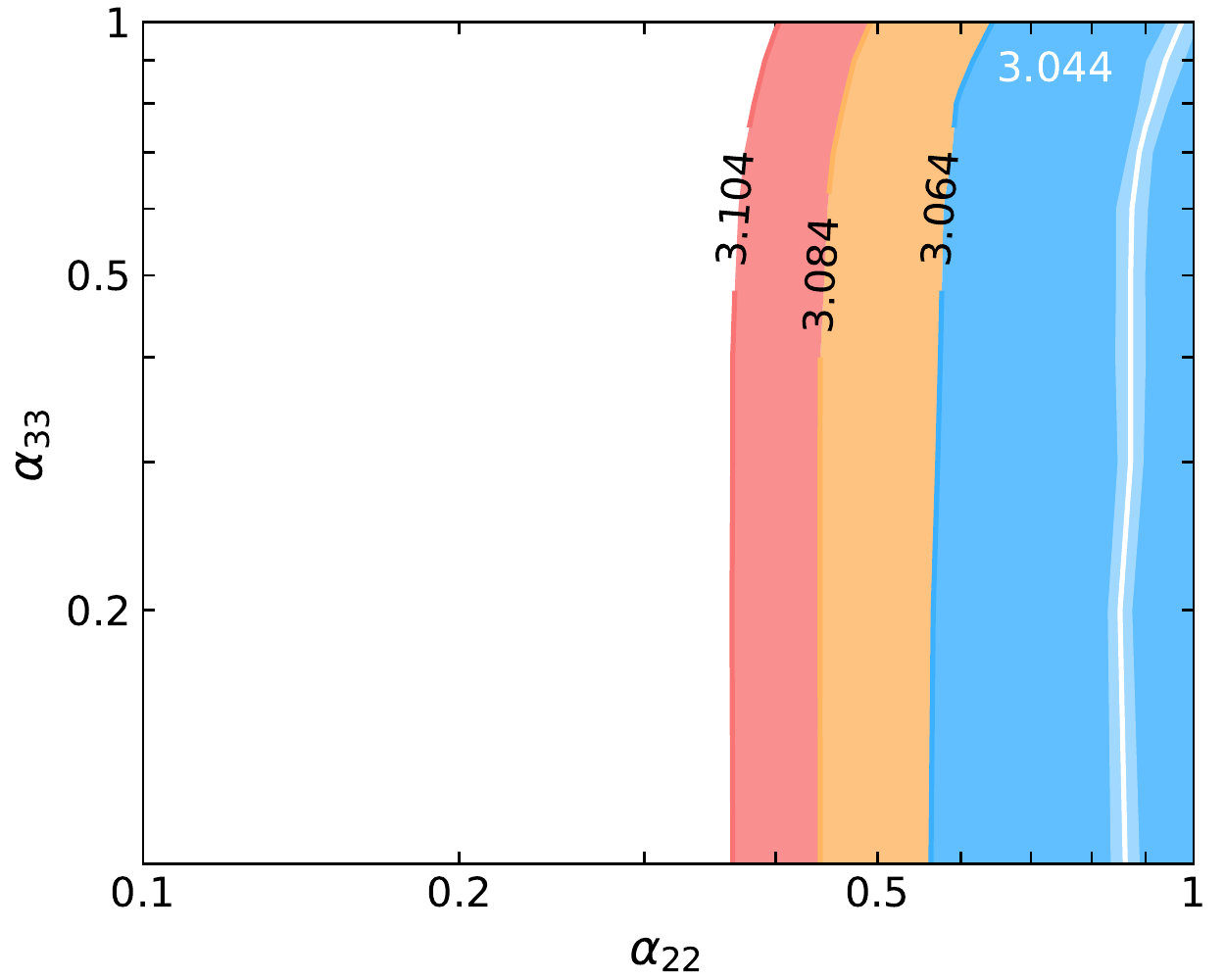}
    \caption{The three panels show the 1$\sigma$, 2$\sigma$ and 3$\sigma$ allowed regions expected from the measurement of \Neff ~when two of the diagonal parameters are allowed to vary simultaneously. The top panel shows the $\alpha_{11} - \alpha_{22}$ plane, whereas the bottom left and bottom right panels correspond to the $\alpha_{11} - \alpha_{33}$ and $\alpha_{22} - \alpha_{33}$ planes, respectively. The isocontour of \Neff =$(N_{\rm eff})_0$=3.044  is shown in white, together with a shaded band indicating the numerical uncertainty, $(N_{\rm eff})_0\pm 0.001$.}
    \label{fig:nu_two_at_a_time}
\end{figure}
Figure \ref{fig:nu_two_at_a_time} depicts the expected sensitivity to the non-unitarity diagonal parameters, projected in two-dimensional contours. The three coloured regions correspond to the 1$\sigma$, 2$\sigma$ and 3$\sigma$ deviations from the standard value for \Neff, shown by the white line. The white shaded band corresponds to an estimate of the numerical error in the standard prediction, given the numerical settings chosen in this work. 

In the top panel, we show how the expected limit on \Neff\ translates into allowed regions in the $\alpha_{11}-\alpha_{22}$ two-dimensional plane. Although these regions are larger than the ones derived from the comparison of the measurements of $G_{\rm F}$ from $\beta$ and muon decays or from neutrino oscillation experiments, they provide independent constraints on the non-unitarity diagonal parameters. The lower panels correspond to the sensitivity in the $\alpha_{11}-\alpha_{33}$ and $\alpha_{22}-\alpha_{33}$ two-dimensional planes. As it was already shown in Figure \ref{fig:one_at_a_time}, values of $\alpha_{33}$ which deviate significantly from unity do not result in a relevant change in \Neff: in particular, for values of $\alpha_{33}< 0.5$, \Neff\ becomes almost independent of $\alpha_{33}$.
The reason for the small dependence of \Neff\ on $\alpha_{33}$ comes from the fact that the former only affects interaction rates involving the $\tau$ neutrino, which have a minor role in the neutrino decoupling process. On the contrary, $\alpha_{11}$ and $\alpha_{22}$ tend to result in a larger value of \Neff\ because of their effect on $G_{\rm F}$.

\subsection{Including off-diagonal parameters}
Allowing for non-zero off-diagonal parameters, $\alpha_{ij} > 0$ with $i < j$, changes the prediction for the effective number of relativistic species, as these parameters modify the structure of neutrino interactions with the plasma as well. In addition, for non-zero $\alpha_{21}$, the relation between the effective Fermi constant measure from muon decay, $G_{\rm F}^\mu$, and the corresponding fundamental constant $G_{\rm F}$ is altered. In what follows, we present the results of adding one extra non-zero off-diagonal parameter to the three diagonal ones, limiting the analysis to real values of $\alpha_{ij}$\footnote{This choice is motivated by the results in \cite{Froustey:2021azz}, where it was shown that the CP phase does not alter the predicted \Neff\ in the case of unitary mixing.}.

\begin{figure}
    \centering
    \includegraphics[width=0.56\textwidth]{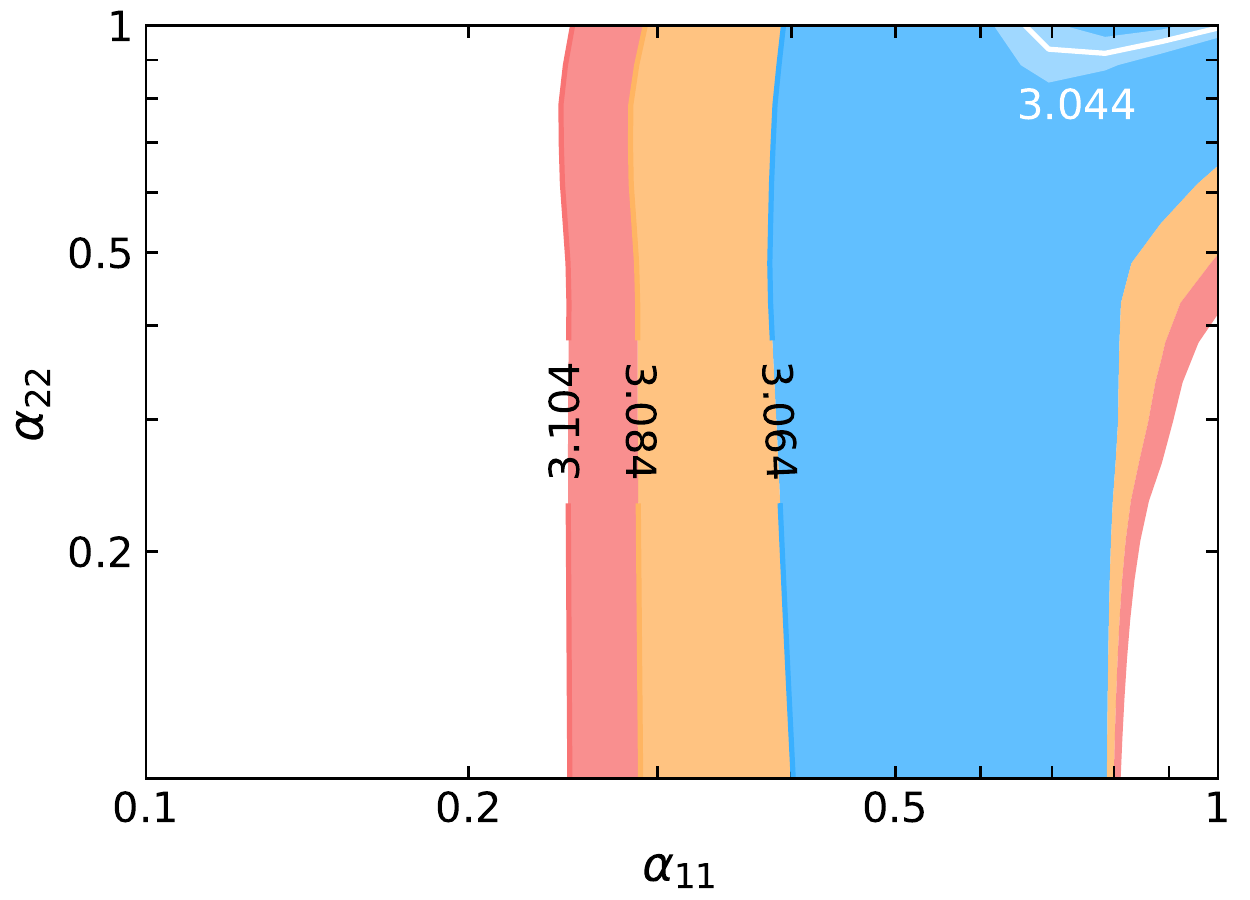}
    \caption{Expected allowed regions at 1$\sigma$, 2$\sigma$ and 3$\sigma$ significance from the measurement of \Neff ~in the $\alpha_{11}-\alpha_{22}$ plane, when $\alpha_{21}$ is allowed to vary while respecting the unitarity condition Eq.~(\ref{eq:NUcond1}) and after marginalizing over  $\alpha_{21}$, see main text for details.}
    \label{fig:3at_a_time}
\end{figure}

We show in Figure \ref{fig:3at_a_time} how non-zero values of $\alpha_{21}$ satisfying Eqs.~(\ref{eq:NUcond1})-(\ref{eq:NUcond3}) enlarge the parameter space allowed in the $\alpha_{11}-\alpha_{22}$ two-dimensional plane  after marginalising over the $\alpha_{21}$ parameter, see the upper left panel of Figure~\ref{fig:nu_two_at_a_time} for comparison.
Notice that non-zero values of the $\alpha_{21}$ parameter can partially compensate the larger value of \Neff\ expected when $\alpha_{11}$ and $\alpha_{22}$ depart from one, thanks to its effect on $G_{\rm F}$.
Conversely, non-zero values of the off-diagonal parameters $\alpha_{31}$ and $\alpha_{32}$ were found to introduce only sub-leading changes in the planes $\alpha_{11}-\alpha_{33}$ and $\alpha_{22}-\alpha_{33}$, respectively. This implies that $\alpha_{31}$ and $\alpha_{32}$ do not reduce the value of \Neff\ but rather increase it, leaving therefore the two-dimensional contours in the $\alpha_{11}-\alpha_{33}$ and $\alpha_{22}-\alpha_{33}$ planes almost unchanged.

\section{Discussion}
\label{sec:discussion}

\subsection{From non-unitarity to non-standard interactions}

In the case in which the deviations from unitarity of the three-neutrino mixing matrix are small, one can write the effective Lagrangian from charged- and neutral- current interactions in terms of the active neutrino states, as
\begin{align}
    \mathcal{L}_{NU} = 
    &  -2\sqrt{2}G_{\rm F} \left[ \sum_{\alpha , \beta}\left((N N^\dagger)_{\alpha e}( N N^\dagger)_{e \beta} +g_L \left[ N N^\dagger\right]^2_{\alpha \beta}\right)\left(\bar{\nu}_\alpha \gamma^\mu P_L \nu_\beta\right)\left(\bar{e}\gamma_\mu P_L e \right) \right. \nonumber \\ & \left.+ g_R\sum_{\alpha , \beta}\left[ N N^\dagger\right]^2_{\alpha \beta}\left(\bar{\nu}_\alpha \gamma^\mu P_L \nu_\beta\right)\left(\bar{e}\gamma_\mu P_R e \right)\right]\, .
    \label{eqn:nu}
\end{align}

This means that, if we take into account that the value of $G_{\rm F}$ measured from muon decay is given by Eq.~(\ref{eq:GFmu}), then one can map the scenario of small deviations from  unitarity to the one of neutrino-electron neutral current non-standard interactions (NSI). If we define the corresponding neutral current Lagrangian with NSI as
\begin{align}
        \mathcal{L}_{SM+NSIe}=& - 2\sqrt{2}\,G_{\rm F}  \left(\overline{\nu}_e \gamma^\mu P_{L} e \right) (\overline{e}\gamma_\mu P_{L} \nu_e) \nonumber\\ &- 2\sqrt{2}\,G_{\rm F} \sum_{X, \alpha}\left[ g_{X}\left(\overline{\nu}_\alpha \gamma^\mu P_L \nu_\alpha \right) (\overline{e}\gamma_\mu P_{X} e) + \sum_{\beta} \varepsilon_{\alpha \beta}^{eX} \left(\overline{\nu}_\alpha \gamma^\mu P_L \nu_\beta \right) (\overline{e}\gamma_\mu P_{X} e)\right],
\label{eq:nsie}
\end{align}
then, the NSI-like parameters read as
\begin{align}
    \varepsilon ^{eL}_{\alpha \beta} = -(\delta_{\beta e}\delta_{\alpha e} +g_L\delta_{\alpha\beta}) + \frac{(N N^\dagger)_{\alpha e}( N N^\dagger)_{e \beta} + g_L\left[ N  N^\dagger\right]^2_{\alpha \beta}}{\sqrt{\alpha^2_{11} (\alpha^2_{22} + |\alpha_{21}|^2)}}~,
\end{align}
and
\begin{align}
    \varepsilon ^{eR}_{\alpha \beta} = -g_R\delta_{\alpha\beta }  + g_R\frac{\left[ N  N^\dagger\right]^2_{\alpha \beta}}{\sqrt{\alpha^2_{11} (\alpha^2_{22} + |\alpha_{21}|^2)}} \, .
\end{align}

The involved matrix elements, ($N N^\dagger)_{\alpha \beta}$, can be computed at leading order \cite{Escrihuela:2016ube} and then, a deeper physical understanding of the results becomes possible. The equivalent NSI parameters resulting from mapping departures from unity in the diagonal parameters are shown in Table \ref{tab:nsi}.
If the interactions between neutrinos and the cosmic plasma become weaker, one expects an earlier decoupling, whereas stronger interactions would keep neutrinos in thermal contact for a longer time, and, consequently, neutrinos could receive more energy from the electron-positron pair annihilation.

\begin{table}[]
\renewcommand*{\arraystretch}{1.6}
    \centering
    \begin{tabular}{|c|l l|}
         \hline NU Parameter  &  \multicolumn{2}{l|}{Effective NSI} \\ \hline \hline
         $\alpha_{11} < $  1 & $\varepsilon^{eL}_{\alpha\beta } $ = &  $ ( 1+g_L)(-1+ \alpha_{11}^3)\delta_{\beta e}\delta_{\alpha e}+ g_L \left(\frac{1}{\alpha_{11}} -1\right)(\delta_ {\alpha \mu}\delta_{\beta \mu} +\delta_ {\alpha \tau}\delta_{\beta \tau})$   \\
         & $\varepsilon^{eR}_{\alpha\beta } = $ &  $ -g_R(1 -\alpha^3_{11})\delta_{\alpha e}\delta_{\beta e} + g_R\left(\frac{1}{\alpha_{11} }-1 \right)(\delta_ {\alpha \mu}\delta_{\beta \mu} +\delta_ {\alpha \tau}\delta_{\beta \tau})$
          \\
         \hline\hline
         $\alpha_{22} < $  1 & $\varepsilon^{eL}_{\alpha\beta } $ = &  $ -(1 +g_L)\left( 1 - \frac{1}{\alpha_{22}}\right)\delta_{\alpha e} \delta_{\beta e} $ \\ & & $-g_L(1-\frac{1}{\alpha_{22}})\delta_{\alpha\tau} \delta_{\beta \tau}- g_L(1 -{\alpha^3_{22}}) \delta_ {\alpha \mu}\delta_{\beta \mu}$   \\
         & $\varepsilon^{eR}_{\alpha\beta } = $ &  $ -g_R\left(1 - \frac{1}{\alpha_{22}}\right)(\delta_{\alpha e}\delta_{\beta e} + \delta_{\alpha \tau }\delta_{\beta \tau} )- g_R(1-\alpha^3_{22}) \delta_ {\alpha \mu}\delta_{\beta \mu} $
        \\
         \hline \hline
         $\alpha_{33} < $  1 & $\varepsilon^{eL}_{\alpha\beta}$ = &  $g_L(-1 + \alpha_{33}^4)\delta_{\alpha\tau}\delta_{\tau\beta}$  \\
         & $\varepsilon^{eR}_{\alpha\beta }  = $ & $ g_R(-1 + \alpha_{33}^4)\delta_{\alpha\tau}\delta_{\tau\beta}$
         \\
         \hline 
    \end{tabular}
    \caption{Mapping between the diagonal non-unitarity (NU) parameters and effective NSI with electrons.}
    \label{tab:nsi}
\end{table}

From Table \ref{tab:nsi}, it is possible to notice that the behaviour of the predictions in Figure~\ref{fig:one_at_a_time} result from the modification of the interactions, which, depending on the values of $\alpha_{11}$ and $\alpha_{22}$, can shift to earlier or later times the neutrino decoupling process.
Conversely, the results obtained for $\alpha_{33}$ are clearly understood from the fact that a value $\alpha_{33} < 1$ reduces the strength of the neutral-current interactions, keeping tau neutrinos in thermal contact with the plasma, and leading to an earlier decoupling.

When two or more parameters are needed to parametrise the non-unitarity of the three-neutrino mixing, the picture becomes more complicated since the different effects can compensate. Nevertheless, one can get an idea of the results obtained when three parameters were allowed to vary simultaneously. For instance, $\alpha_{31} > 0$ and $\alpha_{32} > 0$ result in the appearance of flavour-changing neutral currents, which lead to a delayed decoupling \cite{deSalas:2021aeh}. The scenario of $\alpha_{21}$ is different since it modifies both the structure of interactions and their strength. 

Note that the treatment above is only valid for small departures from unitarity. Other\-wise, the calculations must be performed in the mass basis in order to consider only the kinematically accessible states.

\subsection{Comparison to other limits}
At Earth, several precision and flavour observables can also be used to constrain deviations from unitarity in the three-neutrino mixing matrix. Whereas a detailed global analysis of all the existing constraints is beyond the scope of this work, for completeness, we discuss some of the most restrictive bounds from references \cite{Antusch:2014woa,Antusch:2016brq,Escrihuela:2016ube,Escrihuela:2019mot, Fernandez-Martinez:2016lgt}.

At tree level, the weak mixing angle is related to the Fermi constant by
\begin{align}
    \sin ^2\theta_W \cos^2 \theta_W = \frac{\alpha\pi}{\sqrt{2}G_{\rm F} \,m^2_Z} = \frac{\alpha \pi}{\sqrt{2}G^\mu_F \,m^2_Z}\sqrt{(N N^\dagger)_{ee}(NN^\dagger)_{\mu\mu}}\, ,
\end{align}
where $\alpha$ is the fine structure constant. Requiring the agreement between measurements of the weak mixing angle and its indirect inference using $G_{\rm F}$ allows to set limits on departures from unitarity for the matrix $N$. 

The prediction of weak decay widths of the Z boson also depends on the assumptions about the unitarity of the lepton mixing matrix. At tree level, the decay width to two fermions depends linearly on $G_{\rm F}$, whereas for the invisible decay width,
\begin{align}
    \Gamma_{\rm inv} \propto G_{\rm F} \sum_{i,j = 1,2,3}|(N^\dagger N) _{ij}|^2 \,.
\end{align}

Also at tree level, the mass of the W boson can be expressed in terms of the Fermi constant, $G_{\rm F}$, as
\begin{align}
    m^2_W = \frac{\alpha \pi}{\sqrt{2}G_{\rm F} \sin^2\theta_W}\,.
\end{align}
Again, bounds on departures from unitarity arise from requiring the agreement between direct measurements and the value of $m_W$ derived using $G^\mu_{\rm F}$. 

As mentioned above, the values of $G_{\rm F}$ derived from $\beta$ and muon decays would receive different corrections. As a consequence, limits on the unitarity of the Cabibbo-Kobayashi-Maskawa matrix can be recasted, since
\begin{align}
    \sum_{i=1}^3 |V_{ui}|^2 = \left(\frac{G^\beta_{\rm F}}{G^\mu_{\rm F}}\right)^2 = \frac{1}{\alpha_{22}^2 + |\alpha_{21}|^2}\, .
\end{align}

In the presence of non-unitarity, couplings between leptons and gauge bosons are no longer flavour independent, implying that semileptonic decay rates  can be flavour-dependent. Thus, measurements of pion and kaon decays also restrict deviations from unitarity. One can also consider universality tests in the leptonic sector and W-boson decays. For instance,
\begin{align}
    R ^W_{\alpha \beta} = \frac{\Gamma(W\rightarrow l_\alpha  \bar{\nu}_\alpha ) }{\Gamma(W\rightarrow  l_\beta \bar{\nu}_\beta)} = \sqrt{\frac{(N N^\dagger)_{\alpha\alpha}}{(N N^\dagger)_{\beta\beta}}}\, .
\end{align}

These and other probes have been studied extensively in the literature and provide stringent limits to possible deviations from unitarity of the three-neutrino mixing matrix \cite{Antusch:2016brq,Fernandez-Martinez:2016lgt,Escrihuela:2016ube,Escrihuela:2019mot}. Table \ref{tab:limitsleptons} reports the constraints on the $\alpha_{ij}$ parameters obtained in \cite{Escrihuela:2016ube} when considering constraints from neutrinos and charged leptons. Constraints from neutrinos alone \cite{Forero:2021azc} are complementary albeit not completely competitive. 
\begin{table}[]
    \centering
    \begin{tabular}{|c|c|c|c|c|c|}
    \hline
        $\alpha_{11}$ &  $\alpha_{22}$ &  $\alpha_{33}$ &  $|\alpha_{21}|$ &  $|\alpha_{31}|$ &  $|\alpha_{31}|$   \\\hline
          $>$ 0.9961 & $>$ 0.9990 & $>$ 0.9973 & $<$ 2.6$\cdot 10 ^{-3}$ & $<$ 5.0$\cdot 10 ^{-3}$ & $<$ 2.4$\cdot 10 ^{-3}$ \\ \hline
    \end{tabular}
    \caption{Combined limits on the non-unitarity parameters $\alpha$ from charged lepton and neutrino measurements at 90\% C.L. (considering 6 degrees of freedom), from \cite{Escrihuela:2016ube}.}
    \label{tab:limitsleptons}
\end{table}

\section{Concluding remarks}
\label{sec:conclusion}

In this work, we have considered the impact of non-unitarity in the three-neutrino mixing matrix on the process of neutrino decoupling. The presence of non-unitarity, a smoking-gun low-energy effect for the presence of heavy neutral leptons, alters the structure of both charged and neutral current interactions, thus affecting how neutrinos decouple from the thermal plasma in the early Universe.

In light of the forecasted accuracy expected from forthcoming cosmological surveys, we have derived the deviations on \Neff\ induced by a non-unitarity mixing in the three-neutrino sector. In this regard, future cosmological  measurements can serve as a novel high-redshift tool to provide model-independent and laboratory-complementary constraints on the non-unitarity parameters arising from new physics. Furthermore, they can be used as a cross-check to either falsify (or confirm) neutrino mass-generation scenarios. 

We employ the commonly exploited parameterisation for the non-unitarity parameters in Eq.\ (\ref{eq:parametrisation}). If the diagonal parameters $\alpha_{11}$ and $\alpha_{22}$ are different from unity, they will delay neutrino decoupling, resulting in a larger value of \Neff. The lower limits we obtained here when considering the future CMB-S4 mission combined with large-scale structure observations are $\alpha_{11} > 0.29$ and $\alpha_{22}> 0.50$, both at $95\%$~C.L. Concerning the third non-unitarity diagonal parameter $\alpha_{33}$, it would barely alter the overall picture. 
We have also explored the possible interplay between the diagonal and off-diagonal non-unitarity parameters when either two or three of them are allowed to vary simultaneously. Interestingly, a non-zero value $\alpha_{21}$ relaxes considerably the constraints in the $\alpha_{11}-\alpha_{22}$ plane. 

We conclude here emphasizing the fact that, even if the constraints on the non-unitarity parameters expected from cosmological observations are among the weakest when compared to other terrestrial probes, they provide an independent limit and a consistency test of the neutrino decoupling picture. Furthermore, they allow to set constraints on the putative non-unitary three-neutrino mixing at a completely different epoch in the universe evolution. Therefore, they should not be regarded as competitive limits but as supportive searches.

\acknowledgments
%
SG and OM acknowledge financial support from the European Union's Horizon 2020 research and innovation programme under the Marie Skłodowska-Curie grant agreement No 754496 (project FELLINI) and the ITN grant H2020-MSCA-ITN-2019/860881-HIDDeN,
respectively. In Valencia, this work was supported by the Spanish grants
PID2020-113775GB-I00 and PID2020-113644GB-I00 (MCIN/AEI/10.13039/501100011033), as well as projects PROMETEO/2019/083 and CIPROM/2021/054 (Generalitat Valenciana). PMM is also supported by the grant FPU18/04571 (MICIU). He is grateful for the hospitality of the Particle and Astroparticle Physics 
Division of the \mbox{Max-Planck-Institut} f{\"u}r Kernphysik (Heidelberg), the Centro de Física Teórica de Partículas (Instituto Superior Técnico, Lisbon) and the AstroNu group at Niels Bohr Institute (Copenhagen) during the development of this project.

\appendix
\section{Parametrisation of the non-unitary three-neutrino mixing matrix}
\label{sec:mixing}

The non-unitarity of the three-neutrino mixing matrix can be parametrised by a triangular matrix. The parameters introduced can be related to the full $n\times n$ unitary lepton mixing matrix, $U_{n \times n}$. If we define the following complex rotation matrix,
\begin{align}
    \omega_{ij} = \begin{pmatrix}
    1& 0 & & \cdots &0& \cdots& & &0\\
    0& 1 & & & & & & & \vdots \\
    \vdots & & c_{ij} & \cdots & 0 & \cdots & \eta_{ij} & & \\
    & & \vdots & \ddots & &  &\vdots & \\
    & & 0 & &1& & 0 & &\\
    & & \vdots & & \ddots& &\vdots & &\\
    & & \bar{\eta}_{ij} &\cdots&0& \cdots& c_{ij}&  &\vdots \\
    \vdots & & & & & & & 1 & 0 \\
    0 & & & \cdots & 0 & \cdots& & 0 & 1\end{pmatrix} \, ,
\end{align}
where we use the notation $c_{ij} = \cos \theta_{ij}$, $\eta_{ij} = \sin \theta_{ij}\, e^{- \phi_{ij}}$ and $\bar{\eta}_{ij} = -\sin \theta_{ij} \,e ^{i\phi_{ij}}$, the full lepton mixing matrix can be parametrised as
\begin{align}
    U_{n\times n} = \omega_{n-1 \, n}\, \omega_{n-2 \, n} ...\, \omega_{1\, n} \omega_{n-2 \, n-1}\, \omega_{n-3\, n-1} ...\, \omega_{1\, n-1} ...\, \omega_{23}\,\omega_{13}\,\omega_{12}\, .
\end{align}
The diagonal parameters $\alpha_{ii}$ in Eq.\ \eqref{eq:parametrisation} are real and expressed as \cite{Escrihuela:2015wra,Forero:2021azc}
\begin{align}
    \alpha_{ii} = c_{i\,n}\, c_{i\, n-1}...\, c_{i\, 4} \, ,
\end{align}
and the off-diagonal ones are 
\begin{align}
    \alpha_{21} = \,&c_{2\,n}\,c_{2\, n-1} ...\, c_{25}\,\eta_{24}\,\bar{\eta}_{14} + c_{2\,n}...\, c_{26}\,\eta_{25}\,\bar{\eta}_{15}\,c_{14} + ...+\eta_{2\,n}\,\bar{\eta}_{1n}\,c_{1\,n-1}...\,c_{14} \, ,\\
     \alpha_{31} =\, &c_{3\,n\,}c_{3\, n-1}...\, c_{35}\,\eta_{34}\,c_{24}\,\bar{\eta}_{14} +c_{3\,n}...\,c_{36}\,\eta_{35}\,c_{25}\,\bar{\eta}_{15}\,c_{14}+...+\nonumber\\ & \hspace{7.3cm}+\eta_{3\,n}\,c_{2\,n}\,\bar{\eta}_{1\, n}\,c_{1\, n-1}...\, c_{14}\, ,\\
    \alpha_{32} = \,&c_{3\,n}\,c_{3\,n-1}...\, c_{35}\,\eta_{34}\,\bar{\eta}_{24}+ c_{3\,n}...\,c_{36}\,\eta_{35}\,\bar{\eta}_{25}\,c_{24} + ... +\eta_{3\, n}\,\bar{\eta}_{2\,n}\, c_{2\, n-1} ...\, c_{24}\, .
\end{align}

\section{Collision integrals in the mass basis}
\label{sec:collison}

Following the notation and procedure for integral reduction in \cite{Bennett:2020zkv}, the neutrino-electron collision integral is separated in the terms accounting for scattering and annihilations,
\begin{align}
    \mathcal{I}_{\nu  e}[\varrho (x,y)] = \frac{G^2_F}{(2\pi)^3y^2}\lbrace I_{\nu e}^{\rm scatt} [\varrho(x,y)] + I_{\nu e}^{ann}[\varrho(x,y)]\rbrace\, ,
\end{align}
using the comoving variables defined in Section \ref{sec:cosmo}. The term accounting for scattering reads
\begin{align}
    I_{\nu e}^{\rm scatt} [\varrho(x,y)] = \int {\rm d}y_2 {\rm d}y_3
\sum_{a=L,R}\sum_{b=L,R}
A_{ab}(x,y,y_2,y_3)
F_{\rm sc}^{ab}\left(\varrho^{(1)}, f_e^{(2)}, \varrho^{(3)}, f_e^{(4)}\right)\, ,
\end{align}
whereas the one related to annihilations is
\begin{align}
    I_{\nu e}^{\rm ann} [\varrho(x,y)] = \int {\rm d}y_2 {\rm d}y_3
\sum_{a=L,R}\sum_{b=L,R}
B_{ab}(x,y,y_2,y_3)
F^{\rm ann}_{ab}\left(\varrho^{(1)}, \varrho^{(2)}, f_e^{(3)}, f_e^{(4)}\right)\, .
\end{align}
The exact dependence on the comoving variables of the scattering kernels, $A_{ab}$ and $B_{ab}$, can be found in~\cite{Bennett:2020zkv}. Their expressions do not depend on the basis in which the evolution of the density matrix is computed (mass basis here, flavour basis in~\cite{Bennett:2020zkv}) nor on the strength and structure of the interactions.

The fact that we are working in the mass basis only manifests in the phase-space factors $F^{\rm ann}_{ab}$ and $F^{\rm scatt}_{ab}$, which depend on the matrices $Y_L$ and $Y_R$ defined in Eqs.~(\ref{eqn:YL})-(\ref{eqn:YR}). Although not stated explicitly, the density matrix $\varrho$ is also in the mass basis. Then, the expressions of the factors accounting for the phase-space of scattering and annihilation processes read
\begin{eqnarray}
F^{\rm scatt}_{ab}\left(\varrho^{(1)}, f_e^{(2)}, \varrho^{(3)}, f_e^{(4)}\right)
&=&
f_e^{(4)}(1-f_e^{(2)})\left[Y_a\varrho^{(3)}Y_b(1-\varrho^{(1)})+(1-\varrho^{(1)})Y_b\varrho^{(3)}Y_a\right]
\nonumber\\
&-&
f_e^{(2)}(1-f_e^{(4)})\left[\varrho^{(1)}Y_b(1-\varrho^{(3)})Y_a+Y_a(1-\varrho^{(3)})Y_b\varrho^{(1)}\right]\, ,
\label{eq:F_ab_sc}\\
\nonumber\\
F^{\rm ann}_{ab}\left(\varrho^{(1)}, \varrho^{(2)}, f_e^{(3)}, f_e^{(4)}\right)
&=&
f_e^{(3)}f_e^{(4)}\left[Y_a(1-\varrho^{(2)})Y_b(1-\varrho^{(1)})+(1-\varrho^{(1)})Y_b(1-\varrho^{(2)})Y_a\right]
\nonumber\\
&-&
(1-f_e^{(3)})(1-f_e^{(4)})\left[Y_a\varrho^{(2)}Y_b\varrho^{(1)}+\varrho^{(1)}Y_b\varrho^{(2)}Y_a\right],
\label{eq:F_ab_ann}
\end{eqnarray}
where $a,b = L,R$. These expressions depend both on the electron momentum distribution function, $f_e$, and on the density matrix $\varrho^{(i)}= \varrho(y_i)$, where $y_i$ is the comoving momentum of particle $i$.

\bibliographystyle{JHEP}

\bibliography{bibliography}

\end{document}